\documentclass[12pt]{iopart}
\usepackage{epsfig}
\usepackage{epsf}
\begin{document}
\title{Is Strangeness Chemically Equilibrated?}
\author{Jean Cleymans}
\address{UCT-CERN Research Centre and Department of Physics
University of Cape Town,\\
Rondebosch 7701, South Africa}
\ead{jean.cleymans@uct.ac.za}
\begin{abstract}
Results related to the possible 
chemical  equilibration of hadrons in heavy ion collisions are reviewed.  
Overall
the evidence  is very strong with a few clear and well-documented deviations, 
especially concerning multi-strange hadrons.
Two effects are considered in some detail. Firstly, the neglect of
(possibly an infinite number) of heavy resonances is investigated with the  
help of the Hagedorn model. Secondly, possible deviations from the 
standard statistical distributions are investigated by considering 
in detail results obtained using the Tsallis distribution.
\end{abstract}
\maketitle
\section{Particle Yields}
\begin{figure}[tbh]
\begin{center}
\includegraphics[width=7.5cm]{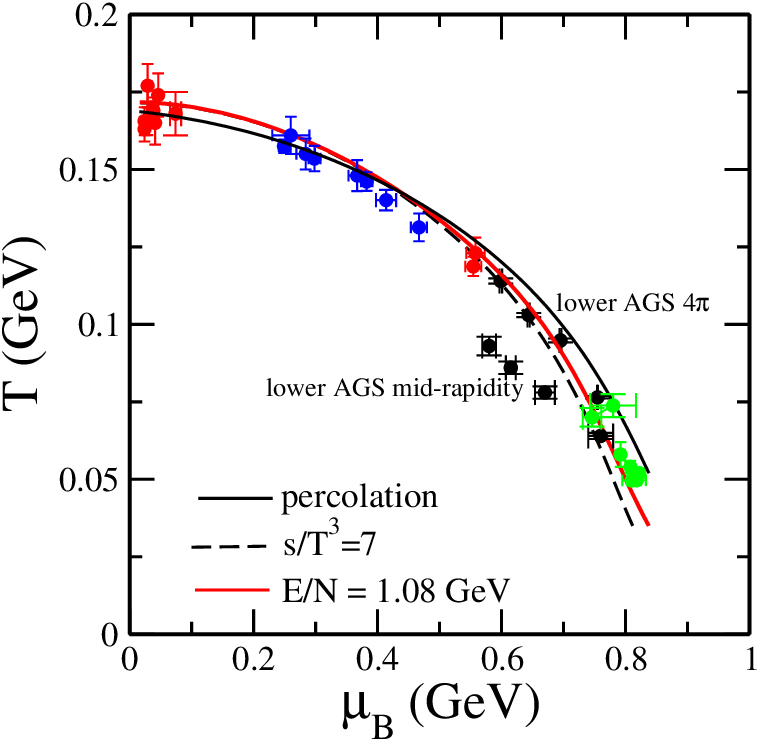}
\label{eovern_2009}
\caption{Values of the freeze-out parameters obtained at beam energies ranging from 1 GeV to 200 GeV}
\end{center}
\end{figure}

\begin{figure}
\begin{center}
\includegraphics[width=7.5cm]{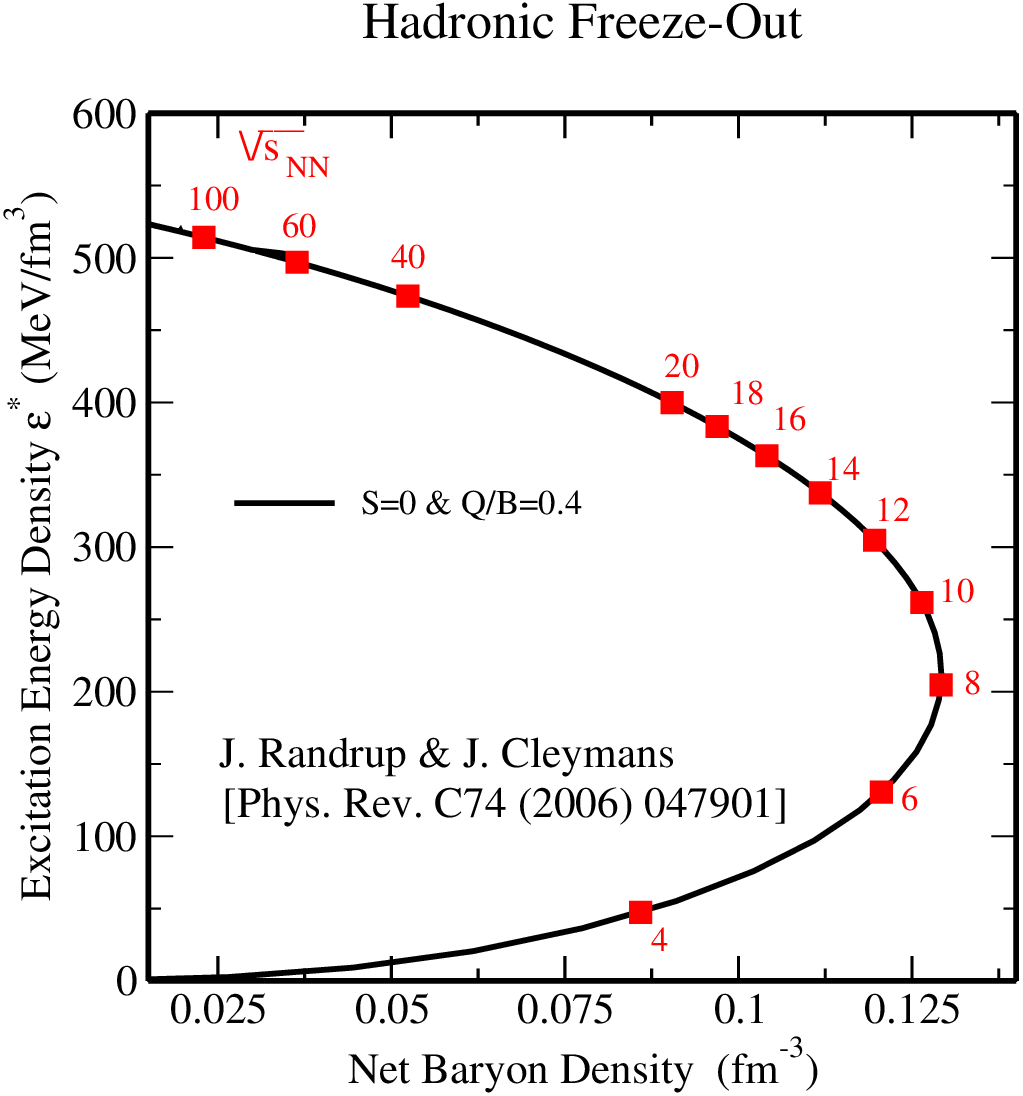}
\caption{The hadronic freeze-out line in the
$\rho_B-\epsilon^*$ 
phase plane as obtained from the values of $\mu_B$ and $T$
 that have been extracted from the experimental data in~\cite{comparison}.
The calculation employs values of $\mu_Q$ and $\mu_S$ 
that ensure $\langle S\rangle=0$ and $\langle Q\rangle=0.4\langle B\rangle$
for each value of $\mu_B$.  
Also indicated are the beam energies (in GeV/N)
for which the particular freeze-out 
conditions are expected at either RHIC or FAIR or NICA. 
}
\label{randrup}
\end{center}
\end{figure}

After  analysing particle multiplicities for two decades
a remarkably simple picture has emerged for the chemical freeze-out 
parameters~\cite{stachel,becattini,comparison}. 
Despite much initial skepticism, the thermal model
has emerged as a  reliable guide for particle multiplicities 
in heavy ion collisions at all collision  energies.
Some of the results, including  analyses from~\cite{davis,jun,hades,fopi},
are summarised in Fig.~\ref{eovern_2009}.
 Most of the points in 
Fig.~\ref{eovern_2009} 
(except obviously the ones at RHIC) refer to integrated ($4\pi$)
yields. A clear discrepancy exists in the lower AGS beam energy region
between the (published) mid-rapidity yields 
and estimates of the $4\pi$ yields. 
The latter tend to give higher 
values for the chemical freeze-out temperature. This will have to be 
resolved by future experiments at e.g. NICA and FAIR.
When the temperature and baryon chemical potential 
are translated to net baryon and energy densities, 
a different, but equivalent, picture emerges shown in Fig.~\ref{randrup}.
This clearly shows the importance in going to the beam energy region
of around 8 - 12 GeV as this corresponds to the highest freeze-out 
baryonic density 
and to a rapid change in thermodynamic 
parameters~\cite{randrup1,randrup2}.

The dependence of $\mu_B$ on the invariant beam energy, $\sqrt{s_{NN}}$, can be 
parameterized as~\cite{comparison}
$$
\mu_B(\sqrt{s_{NN}}) = \frac{1.308~\mathrm{GeV}}{1 + 0.273~{\mathrm{GeV}}^{-1}\sqrt{s_{NN}}}.
$$
Similar dependences
have been obtained by other groups~\cite{stachel,becattini}. 
and are  consistent with the above.  
This predicts at LHC $\mu_B\approx 1$~MeV.

\begin{figure}
\begin{center}
\includegraphics[width=7.5cm]{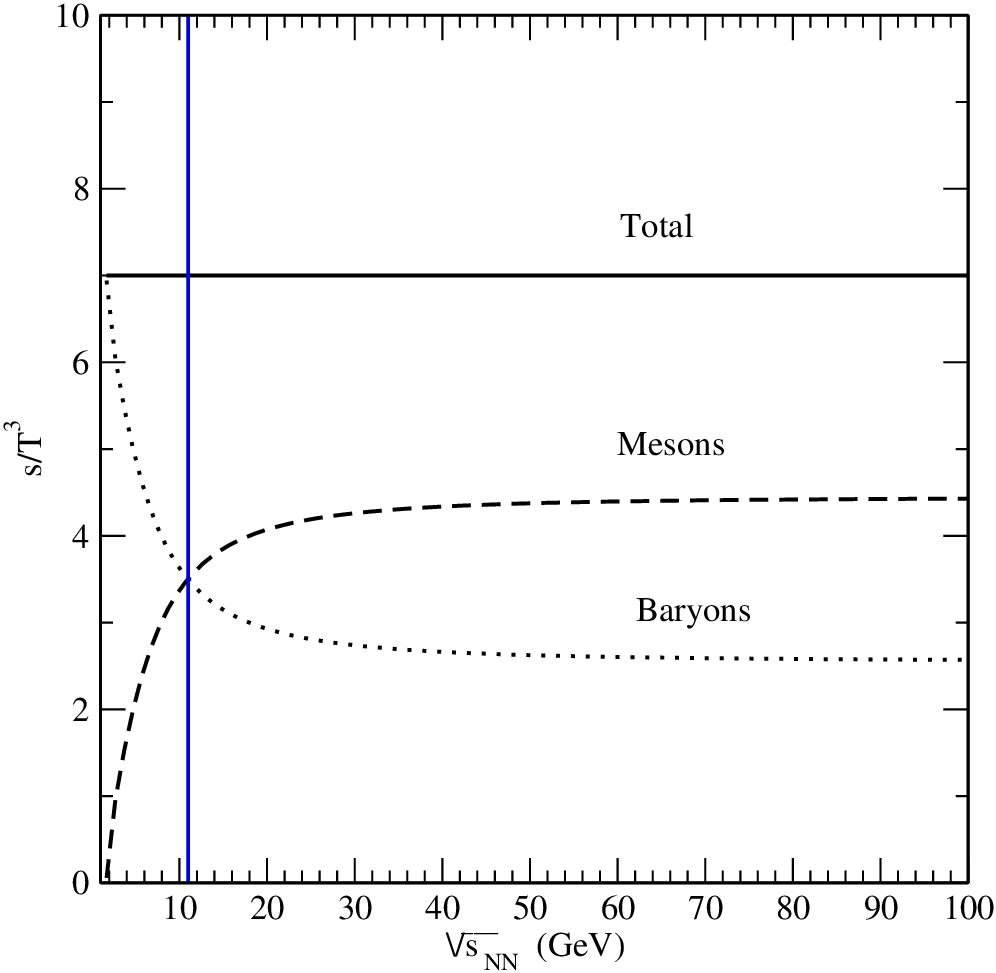}
\caption{Values of entropy density divided by $T^3$  following the chemical freeze-out 
values.}
\label{st3}
\end{center}
\end{figure}

To analyze the changes around 10 GeV use can
be made of the 
entropy density, $s$, divided by $T^3$ which has been shown to reproduce the 
freeze-out curve~\cite{comparison} very well. This allows for a separation into 
baryonic and mesonic components, shown in Fig.~\ref{st3}, it can be 
seen that
mesons  dominate the 
chemical freeze-out from about $\sqrt{s_{NN}}\approx$ 10 GeV onwards.
\begin{figure}
\begin{center}
\label{transition}
\includegraphics[width=7.5cm]{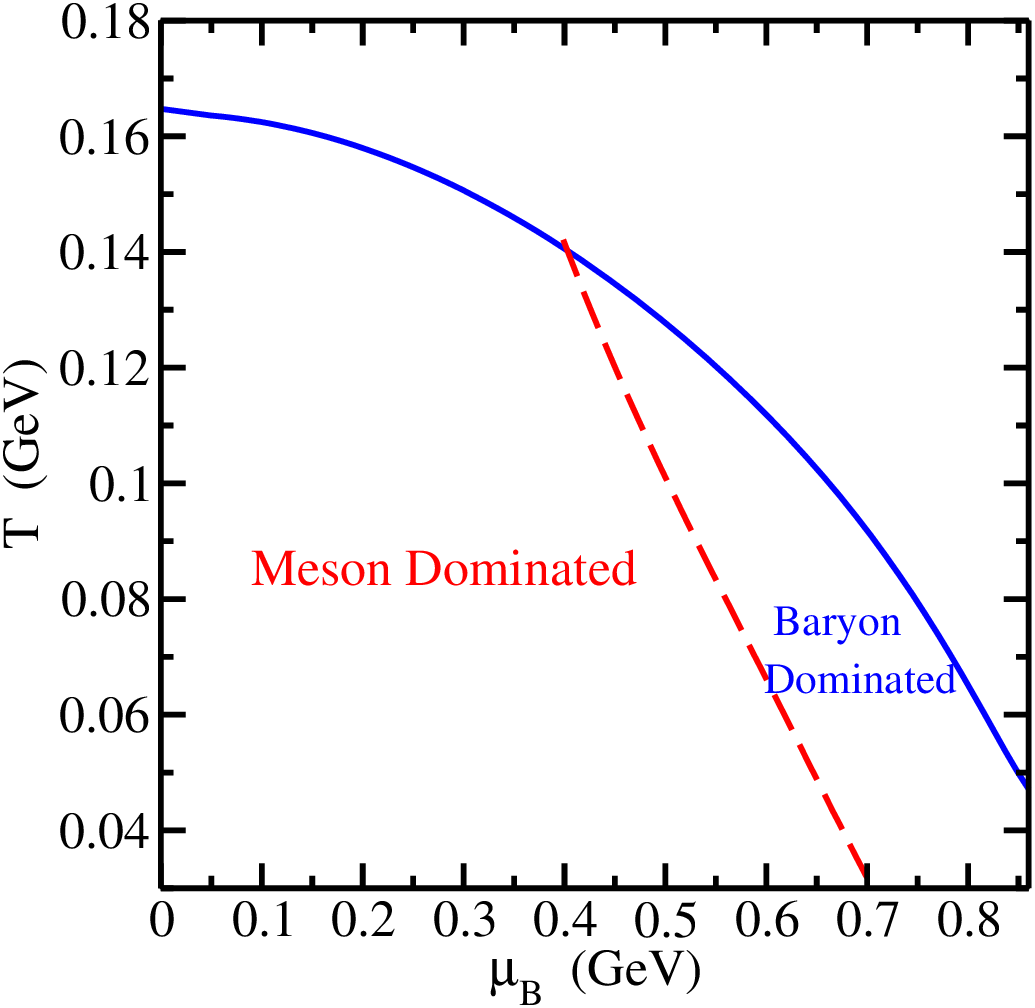}
\caption{Regions in the $T-\mu_B$ plane where baryons or mesons dominate
as indicated.}
\end{center}
\end{figure}
\section{Resonance Gas and Hagedorn Spectrum}
It is possible to compare analytically the resonance gas which uses
 a finite number of resonances up to some maximum mass, 
with a Hagedorn gas which contains an infinite number of resonances with an
exponentially rising number of resonances as the mass increases. It is well known that 
at some point the Hagedorn resonance gas will show a 
divergence when the temperature reaches the Hagedorn 
value~\cite{sound,gupta,jaki,jaki2}.\\
The speed of sound is given by
$$
c_s^2 = \left.{\partial P\over\partial \epsilon}\right|_{s/n}
$$
where the derivative is taken at a fixed entropy per particle.\\
For an ideal Boltzmann gas of identical scalar particles of mass 
$m_0$  and three charge states (``pions'') contained in a volume $V$, 
the grand partition function is defined as
$$
Z(T,V)~=~{{\sum}_N}{\frac{1}{N!}}{\Bigg[\frac{3~V}{(2\pi)^3}{\int}d^3p
\exp\{-\sqrt{p^2 + {m_0}^2}/T\}\Bigg]}^{N}.
$$
This expression can be evaluated, giving
$$
\ln Z(T,V)~=~3{\frac{VT{m_0}^2}{2{\pi}^2}}K_2(m_0/T),
\label{logpart}
$$
The speed of sound in this gas is given by
\begin{equation}
\frac{1}{c^2_s}-3~=~
\frac{{m_0}^2K_2(m_0/T)}{4T^2K_2(m_0/T)+m_0TK_1(m_0/T)} 
\label{soundspeedideal}
\end{equation}
Now extend this to an ideal Boltzmann gas of resonances~\cite{sound},
described by an exponentially increasing mass spectrum of the 
Hagedorn form 
\begin{equation}
\rho(m)~=3 \delta(m-m_0)~ + ~Am^{-4}\exp\{m/T_c\}~\theta(m-2m_0). 
\label{massspectrum}
\end{equation}
The $\theta$-function
assures that the resonance spectrum starts above the two-pion
threshold.  The grand partition function is now given by
\begin{eqnarray}
\ln Z(T,V)~&=&~{VT\over 2\pi^2}3 m_0^2 K_2(m_0/T)\\
&+&{VT\over 2\pi^2}
A{\int}^{\infty}_{\!2m_0}dm~m^{-2}~\exp\{m/T_c\}~K_2(m/T)   ,
\label{logpartmass}
\end{eqnarray}
and the  speed of sound by
\begin{equation}
\frac{1}{c^2_s}~-3~=
$$
$${3 m_0^4K_2(m_0/T)~+~
A \int^{\infty}_{2m_0}dm~\exp\{m/T_c\}~K_2(m/T)
\over
2\pi^2 T s_0~+~
A~T{\int}^{\infty}_{2m_0}dm~ {m}^{-2}~\exp\{m/T_c\}~\left[4T K_2(m/T)~+~
m K_1(m/T)\right]} .
\label{soundspeedres}
\end{equation}
The
second term in the numerator diverges as $T\to T_c$, which in turn
causes the speed of sound to vanish there~\cite{sound,gupta,jaki}.
\begin{figure}
\begin{center}
\includegraphics[width=7.5cm]{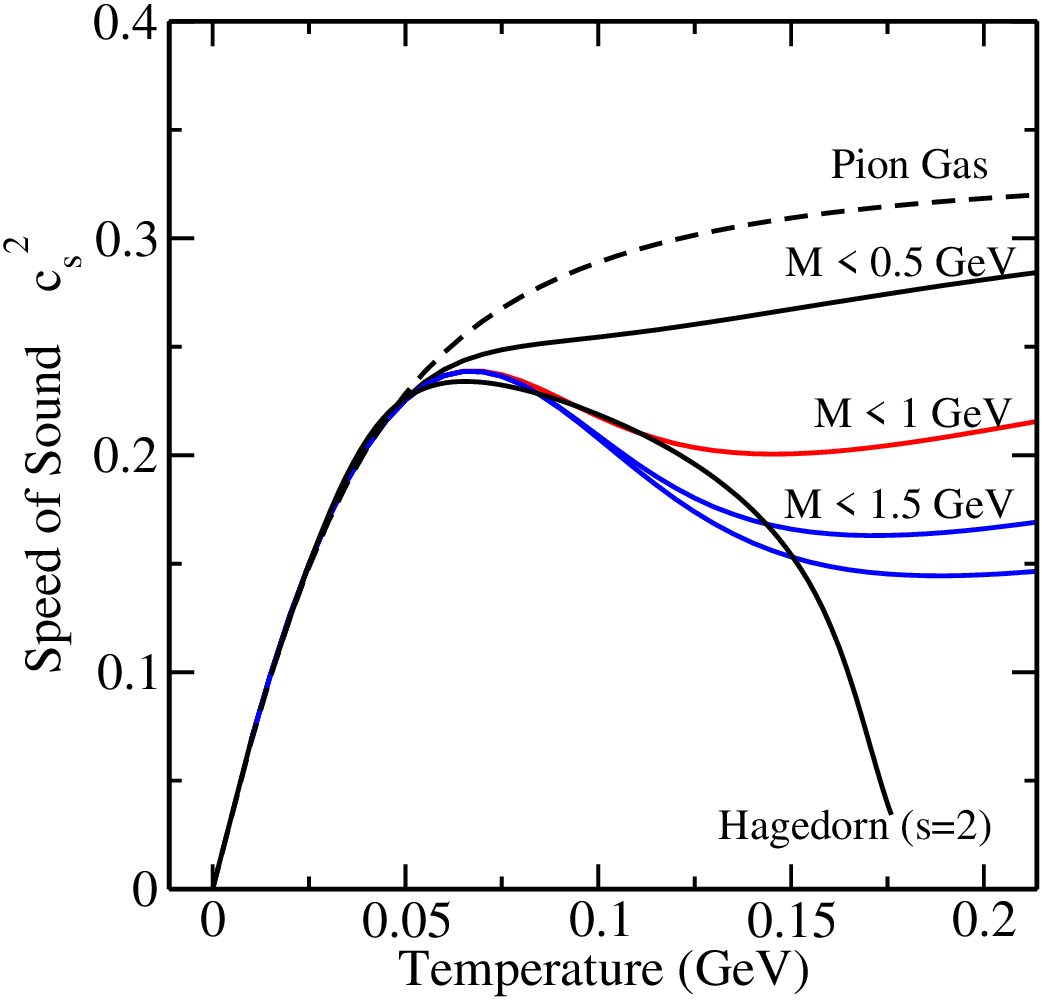}
\label{sound_hg}
\caption{Speed of sound calculated using the thermal model, following the 
values of the chemical freeze-out curve but with different contributions to the resonance gas
determined by the masses of the resonances.} 
\end{center}
\end{figure}
At low temperatures there is almost no difference between a Hagedorn gas and a 
thermal model containing only a limited number of resonances, however at higher
temperatures the calculated speeds of sound become very different as shown
 in Fig.~\ref{sound_hg}.
It is thus always necessary to check if results 
obtained in the thermal model are stable against the addition of 
a Hagedorn-type of mass spectrum. Fortunately, for many quantities of interest
the answer is yes~\cite{gupta}.
\section{Non-extensive Tsallis statistics}
For the Tsallis distribution~\cite{tsallis}, one replaces the standard 
expression
for the entropy based on 
\begin{equation}
S = -\sum_i p_i \ln p_i ,
\end{equation}
with
\begin{equation}
S_T = { 1 -  \sum_i p_i^q \over q -1 } .
\end{equation}
This introduces a new variable $q$, often referred to as the Tsallis parameter.
In the limit where this parameter goes to 1 one recovers the Boltzmann entropy
\begin{equation}
\lim_{q\rightarrow 1}S_T = S
\end{equation}
The physical interpretation of the Tsallis parameter is not obvious, we will
subscribe here to the one presented in Ref.~\cite{wilk}.

We have repeated the analysis for particle yields using the 
Tsallis distribution.
The particle densities of particle yields were calculated 
using~\cite{uct_budapest}:
\begin{equation}
n^T(E) = \left[ 1 + (q-1){E-\mu\over T}\right]^{-1/(q-1)}
\end{equation}
A very interesting application of this distribution 
to the transverse momentum distribution observed in heavy ion collisions 
has been presented
 at this conference  by the STAR collaboration~\cite{sorensen}.
In the limit where the parameter $q$ tends to 1 one recovers the Boltzmann distribution:
\begin{equation}
\lim_{q\rightarrow 1}n^T(E) =  \exp\left(-{E-\mu\over T}\right)
\end{equation}
\begin{figure}
\begin{center}
\includegraphics[width=7.5cm]{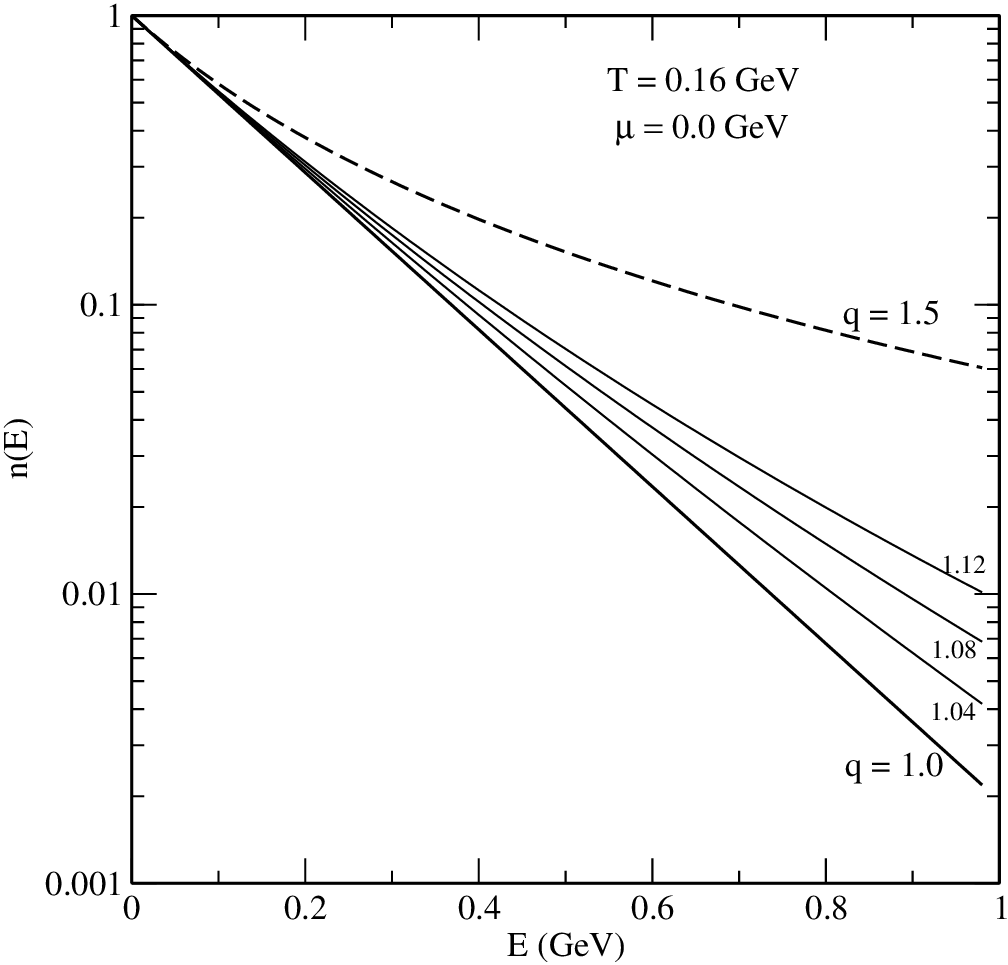}
\label{tsallis_boltzmann}
\caption{Comparison between the Boltzmann and Tsallis distributions.}
\end{center}
\end{figure}
A comparison between the two distributions is shown in 
Fig.~\ref{tsallis_boltzmann}
Clearly, at some value of $q$, an integral over a Tsallis distribution will no longer 
give a convergent result.

A possible interpretation of the  Tsallis Parameter $q$
has been presented in~\cite{wilk}.
One starts by rewriting the Tsallis distribution as a superposition of
Boltzmann distributions with different temperatures, this is possible using a 
distribution function $g$:
\begin{eqnarray}
f(E) &=& \left(1+\left(q-1\right)
\left({E-\mu \over T}\right)\right)^{-1/(q-1)} ,\\\nonumber
&=& \int d\left({1\over T_B}\right)e^{-\left(E-\mu_B\right)/T_B} 
g\left({1\over T_B}\right)  .
\end{eqnarray}
The precise form of the function $g$ has been given  in Ref.~\cite{wilk}.
The average temperature is then given by the $T$ parameter appearing in the Tsallis distribution:
\begin{eqnarray}
\left<{1\over T_B}\right> &=& 
\int  d\left({1\over T_B}\right) \left({1\over T_B}\right)f\left({1\over T_B}\right)\\
&=& {1\over T}
\end{eqnarray}
and the Tsallis parameter $q$  is the deviation around this average Boltzmann temperature,
\begin{equation}
{\left<\left({1\over T_B}\right)^2\right>
- \left<{1\over T_B}\right>^2 
\over
\left<{1\over T_B}\right>^2}
= q - 1
\end{equation}
Thus in the limit where $q$ goes to one, this goes to zero.

\begin{figure}
\begin{center}
\includegraphics[width=7.5cm]{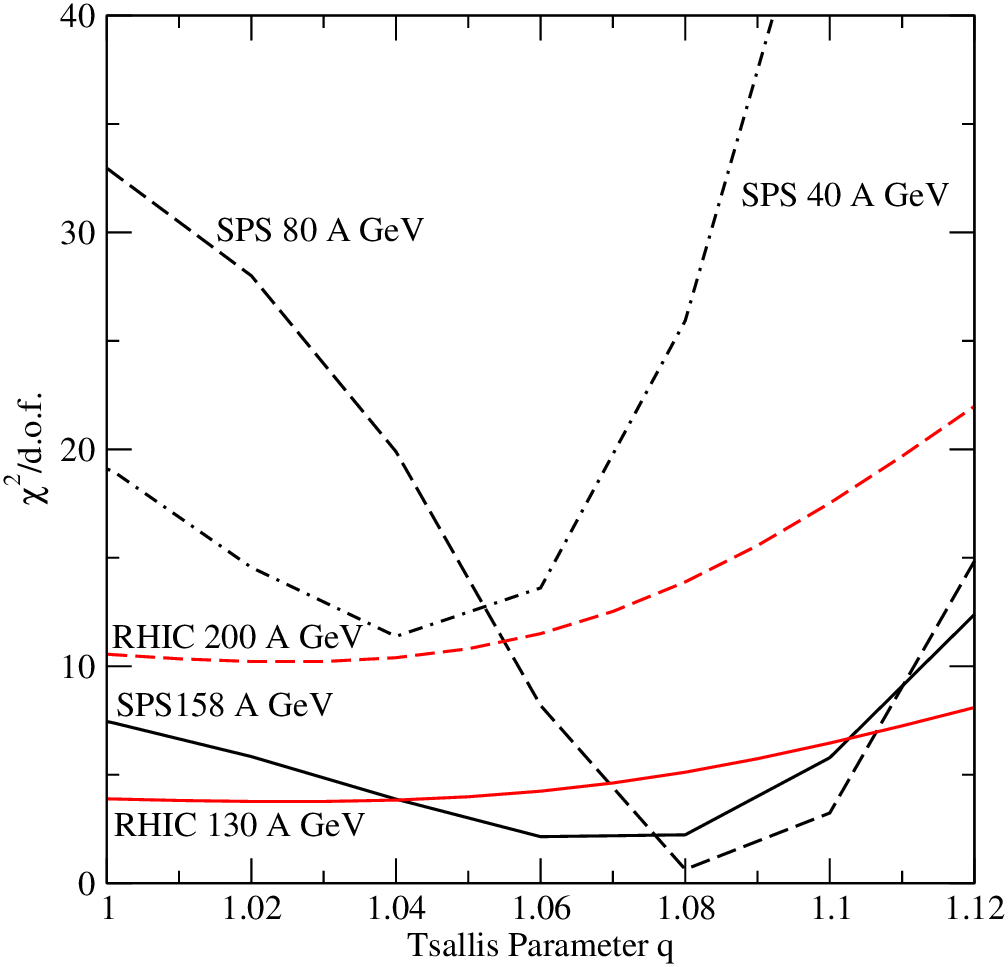}
\label{chi}
\caption{The $\chi^2$/d.o.f. of the fits as a function of the Tsallis
parameter $q$.}
\end{center}
\end{figure}
\begin{figure}
\begin{center}
\includegraphics[width=7.5cm]{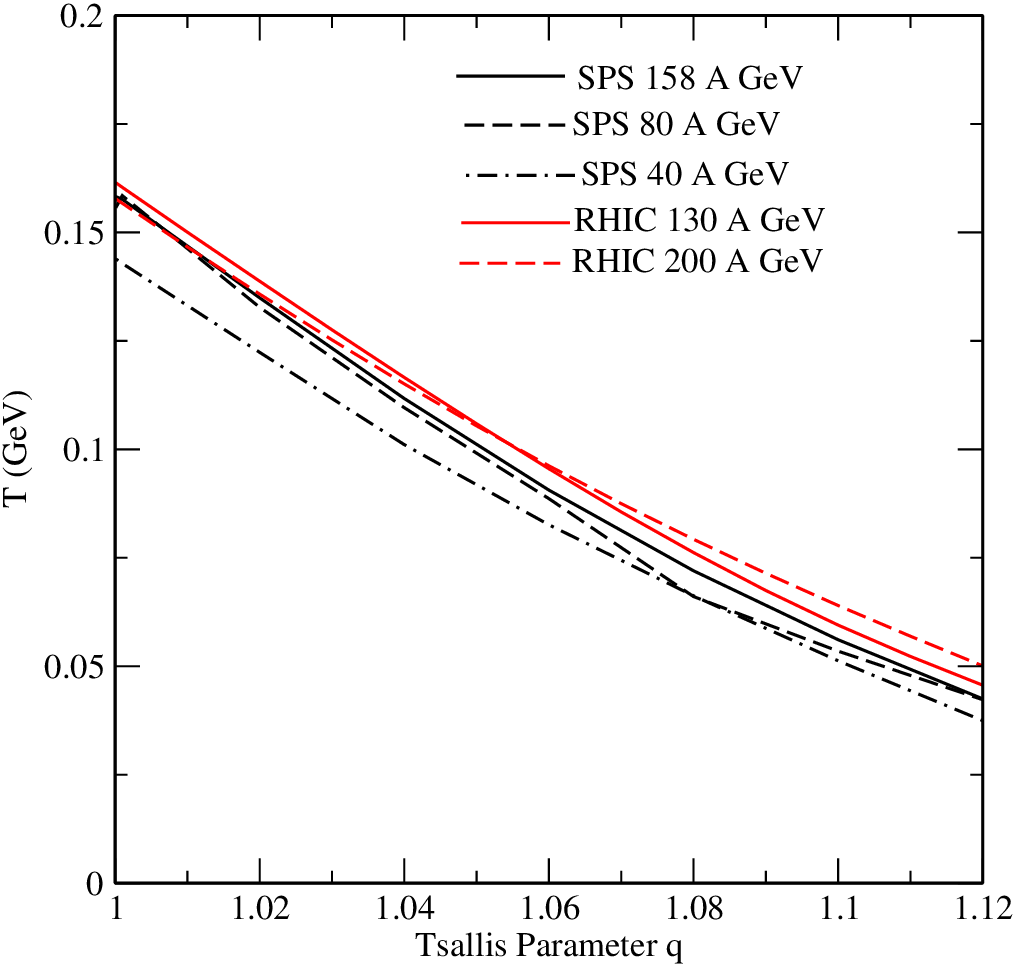}
\caption{The freeze-out temperature as a function of the Tsallis
parameter $q$.}
\label{tsallis-temp}
\end{center}
\end{figure}

The resulting value of  $\chi^2$ shows an interesting 
dependence on the parameter $q$, as shown in Fig.~\ref{chi}.
It must be added that most of the thermodynamic parameters show a very strong dependence on the
parameter $q$ which necessitates a complete review of the physical picture 
behind chemical freeze-out. 
The temperature is shown in Fig.~\ref{tsallis-temp}.
Other variables like the volume and the  chemical potential also 
show a strong variation
with the Tsallis parameter $q$~\cite{uct_budapest}.
\subsection{Acknowledgments}
The author acknowledges many stimulating discussions 
with Tamas Biro, Dan Cebra, Ingrid Kraus, Peter Levai, Helmut Oeschler, Krzysztof Redlich,
Helmut Satz and Spencer Wheaton. 

\end{document}